\documentclass[english,aps,preprint]{revtex4}
\usepackage[T1]{fontenc}
\usepackage[latin9]{inputenc}
\makeatletter
\@ifundefined{textcolor}{}
{%
 \definecolor{BLACK}{gray}{0}
 \definecolor{WHITE}{gray}{1}
 \definecolor{RED}{rgb}{1,0,0}
 \definecolor{GREEN}{rgb}{0,1,0}
 \definecolor{BLUE}{rgb}{0,0,1}
 \definecolor{CYAN}{cmyk}{1,0,0,0}
 \definecolor{MAGENTA}{cmyk}{0,1,0,0}
 \definecolor{YELLOW}{cmyk}{0,0,1,0}
}

\usepackage{slashed}

\makeatother

\usepackage{babel}
\begin{document}

\title{Multigluon tree amplitudes with a pair of massive fermions }

\author{Jia-Hui Huang}

\thanks{corresponding author}

\email{huangjh19@gmail.com}

\affiliation{Center of Mathematical Science, Zhejiang University, Hangzhou, 310027,
PR China}

\affiliation{Zhejiang Institute of Modern Physics, Zhejiang University, Hangzhou,
310027, PR China}

\author{Weijian Wang}

\email{wjnwang96@gmail.com}

\affiliation{Department of Physics, Zhejiang University, Hangzhou, P.R.China}
\begin{abstract}
We consider the calculation of $n$-point multigluon tree amplitudes
with a pair of massive fermions in QCD. We give the explicit transformation
rules of this kind of massive fermion-pair amplitudes with respect
to different reference momenta and check the correctness of them by
SUSY Ward identities. Using these rules and onshell BCFW recursion
relation, we calculate the analytic results of several $n$-point
multigluon amplitudes. 
\end{abstract}
\maketitle

\section{introduction}

Scattering amplitudes are important from both theoretical and experimental
points of view. Traditionally we use Feynman diagrams to calculate
the scattering amplitudes in field theories. During the past several
years, motivated by string theory \cite{TwistorWitten:2003nn}, new
efficient methods for tree level amplitudes have been suggested. The
offshell CSW rule \cite{Cachazo:2004kj} suggests that the tree amplitudes
in gauge theories can be constructed by offshell continued MHV vertices
\cite{mhv}. Then Britto, , Cachazo, Feng and Witten (BCFW) \cite{BCF,BCFW,Feng:2011gc}
gave an onshell recursion relation where the higher point amplitudes
can be obtained by lower point onshell amplitudes whose momenta are
properly shifted to complex space. 

The BCFW recursion relations have been extensively used to calculate
tree level in various massless gauge theories \cite{Luo:2005rx,Luo:2005my,Britto:2005dg,Duhr:2006iq,de Florian:2006vu,de Florian:2006ek,Dinsdale:2006sq,Badger:2010eq}.
Its extension to gravity amplitudes has also been considered \cite{Bedford2005,Cachazo05,Benincasa:2007qj,Elvang:2007sg,Drummond:2009ge,BjerrumBohr:2010ta,BjerrumBohr:2010yc,He:2010ab,Sondergaard:2011iv}.
Those applications are all related to massless external particles.
But in fact, all the matter particles and weak bosons in standard
model are massive. Furthermore, massive amplitudes are more important
in higher energy physics experiments such as LHC experiments for processes
with top quarks, Higgs particles and possible supersymmetric particles.
So it is important to consider the amplitudes with massive external
particles. In \cite{Dixon:2004za,Badger:2004ty,Badger:2006us,Bern:2004ba},
the amplitudes with one external massive gauge bosons or Higgs bosons
have been discussed. Multigluon amplitudes with pairs of massive scalars
or quarks have been studied in \cite{Badger2005,Badger2006,Forde:2005ue,Ozeren:2006ft,Schwinn:2007ee,Feng:2011twa,Craig:2011ws,Chen:2011sba,Boels:2011zz}.
An excellent and compact expression for multigluon amplitudes with
a pair of massive scalars or quarks and any number of plus helicity
gluons has been found in \cite{Ferrario:2006np} by using an off-shell
recursive methods \cite{BG offshell recursive relation} and the BCFW
relations. In \cite{schwinn0602}, the authors use some supersymmetric
Ward identities to relate a compact expression for multigluon helicity
amplitudes involving a pair of massive quarks to amplitudes with massive
scalars. A thorough discussion of BCFW onshell recursion relation
for amplitudes with massive external particles are given in \cite{Schwinn:2007ee}
where the shifted momenta can be massless or massive. The authors
also use BCFW recursion relation to obtain compact expressions for
multigluon amplitudes involving a pair of massive quarks or scalars
and one minus gluon helicity adjacent to fermions. Multigluon amplitudes
with a pair of massive scalar and one minus helicity gluon not adjacent
to scalars are calculated by a different way \cite{Forde:2005ue}.
All helicity amplitudes with a pair of massive quarks are calculated
in \cite{hall} up to six external particles. 

In this paper, we use the notation and convention in \cite{Schwinn:2007ee}
and explore the calculation of several $n$-point helicity amplitudes
with a pair of massive quarks and one minus helicity gluon (fermion-pair
amplitudes). The massive amplitudes depend on reference momenta which
define the helicity of the massive fermions. One can relate the fermion-pair
amplitudes with respect to different reference momenta. We first give
the explicit form of the transformation rules and check the correctness
of them by SUSY Ward identities. Then by shifting the momenta of a
massive particle and a gluon and using the BCFW recursion relation,
we get the analytic expressions of several $n$-point fermion-pair
amplitudes. In section 2, we review the spinor formalism of massive
fermions and obtain the transformation rules for fermion-pair amplitudes
defined on different reference momenta. In section 3, we calculate
several multigluon amplitudes with a pair of fermions. Section 4 is
for the summary.

\section{spinor formalism of massive fermions}

We briefly give our notation and convention of spinor helicity formalism.
For massless fermions, particles and antiparticles both have definite
helicities. Their corresponding spinor states are $u(p,\pm),v(p,\pm)$,
which can be denoted as follows\cite{Schwinn:2007ee}:

\begin{equation}
u(p,\pm)=|p\mp\rangle,v(p,\pm)=|p\mp\rangle.
\end{equation}
For the conjugate states, similar notations are 
\begin{equation}
\bar{u}(p,\pm)=\langle\pm p|,\bar{v}(p,\pm)=\langle\pm p|.
\end{equation}
A massless momentum $q^{\mu}$ can be written in spinor form
\begin{equation}
q^{\mu}=\frac{1}{2}\langle-q|\gamma^{\mu}|q-\rangle=\frac{1}{2}\langle q|\gamma^{\mu}|q]=\frac{1}{2}\langle+q|\gamma^{\mu}|q+\rangle=\frac{1}{2}[q|\gamma^{\mu}|q\rangle.
\end{equation}
The scalar product of two massless momenta $p,q$ can be obtained
as 
\begin{equation}
2p\cdot q=\langle pq\rangle[qp].
\end{equation}

For a massive momentum $K^{2}=m^{2}$, one can always split it into
two massless momenta by introducing a reference massless momentum
$q$,
\begin{equation}
K=K^{\flat}+\frac{K^{2}}{2K\cdot q}q=K^{\flat}+\frac{m^{2}}{2K\cdot q}q,
\end{equation}
where $(K^{\flat})^{2}=0$.

Massive fermions are not helicity eigenstates. Their helicities are
frame dependent and we can introduce a null reference momentum $q$
to define their helicity states. In this formalism, the massive fermions
and anti-fermions states with momentum $p^{2}=m^{2}$ are 
\begin{eqnarray}
u(p,\pm) & = & \frac{1}{\langle p^{\flat}\mp|q\pm\rangle}(\slashed{p}+m)|q\pm\rangle,\\
v(p,\pm) & = & \frac{1}{\langle p^{\flat}\mp|q\pm\rangle}(\slashed{p}-m)|q\pm\rangle,\\
\bar{u}(p,\pm) & = & \langle q\mp|\frac{1}{\langle p^{\flat}\mp|q\pm\rangle}(\slashed{p}+m),\\
\bar{v}(p,\pm) & = & \langle q\mp|\frac{1}{\langle p^{\flat}\mp|q\pm\rangle}(\slashed{p}-m).
\end{eqnarray}
In the above, $p=p^{\flat}+\frac{m^{2}}{2p\cdot q}q$. We can also
define massless states $|p^{\flat}\pm\rangle=\frac{1}{\langle p^{\flat}\pm|q\mp\rangle}\slashed{p}|q\mp\rangle$
and rewrite the massive fermion states as
\begin{equation}
u(p,\pm)=|p^{\flat}\mp\rangle+\frac{m}{\langle p^{\flat}\mp|q\pm\rangle}|q\pm\rangle.
\end{equation}
We can obtain similar forms for all other states. It is easy to see
that we have a smooth massless limit. 

Because the massive fermionic helicity states depend on reference
momenta, then the amplitudes with massive external fermions should
also depend on reference momenta. But we can relate fermionic spinor
states with one reference momentum to these with another reference
momentum\cite{Schwinn:2007ee}. Let $q,\tilde{q}$ be two light-like
reference momenta, we have the following relation between the spinor
states corresponding to the two reference momenta, 
\begin{equation}
\left(\begin{array}{c}
\bar{u}(+)\\
\bar{u}(-)
\end{array}\right)_{\tilde{q}}=\left(\begin{array}{cc}
c_{11} & c_{12}\\
c_{21} & c_{22}
\end{array}\right)\left(\begin{array}{c}
\bar{u}(+)\\
\bar{u}(-)
\end{array}\right)_{q},\label{eq:spinor relation}
\end{equation}
where
\begin{equation}
c_{11}=\frac{\langle\tilde{q}|\slashed{p}|q\rangle}{\langle\tilde{q}\tilde{p}^{\flat}\rangle[p^{\flat}q]},c_{12}=\frac{m\langle\tilde{q}q\rangle}{\langle\tilde{q}\tilde{p}^{\flat}\rangle\langle p^{\flat}q\rangle},c_{21}=\frac{m[\tilde{q}q]}{[\tilde{q}\tilde{p}^{\flat}][p^{\flat}q]},c_{22}=\frac{[\tilde{q}|\slashed{p}|q\rangle}{[\tilde{q}\tilde{p}^{\flat}]\langle p^{\flat}q\rangle}.\label{eq:spinor c xishu}
\end{equation}
In the above equations, $\tilde{p}^{\flat}(p^{\flat})$ means splitting
$p$ with respect to $\tilde{q}(q)$. Similar relations can be obtained
for other massive fermionic states.

In this paper, we consider $n$-point amplitudes with gluons and a
pair of massive fermions $A(1_{Q},2,3,\cdots,n-1,n_{\bar{Q}})$. In
the calculation, we choose the same reference momentum for the external
massive fermions. An amplitude with reference momentum $q$ for the
external fermions is denoted by $A_{q}(1_{Q},2,3,\cdots,n-1,n_{\bar{Q}})$
. In contrast to amplitudes with massless fermions, there are both
helicity-conserving and helicity-flipping amplitudes for massive amplitudes.
With the same external gluons, there are four different helicity amplitudes
$A_{q}(1_{Q}^{+},2,3,\cdots,n-1,n_{\bar{Q}}^{+})$, $A_{q}(1_{Q}^{+},2,3,\cdots,n-1,n_{\bar{Q}}^{-})$,
$A_{q}(1_{Q}^{-},2,3,\cdots,n-1,n_{\bar{Q}}^{+})$, $A_{q}(1_{Q}^{-},2,3,\cdots,n-1,n_{\bar{Q}}^{-})$.
Similar to the relation (\ref{eq:spinor relation}), we can get relations
between amplitudes with respect to different reference momenta, 
\begin{equation}
\left(\begin{array}{c}
A_{\tilde{q}}^{++}\\
A_{\tilde{q}}^{+-}\\
A_{\tilde{q}}^{-+}\\
A_{\tilde{q}}^{--}
\end{array}\right)=\left(\begin{array}{cccc}
C_{11}^{1}C_{11}^{n} & C_{11}^{1}C_{12}^{n} & C_{12}^{1}C_{11}^{n} & C_{12}^{1}C_{12}^{n}\\
C_{11}^{1}C_{21}^{n} & C_{11}^{1}C_{22}^{n} & C_{12}^{1}C_{21}^{n} & C_{12}^{1}C_{22}^{n}\\
C_{21}^{1}C_{11}^{n} & C_{21}^{1}C_{12}^{n} & C_{22}^{1}C_{11}^{n} & C_{22}^{1}C_{12}^{n}\\
C_{21}^{1}C_{21}^{n} & C_{21}^{1}C_{22}^{n} & C_{22}^{1}C_{21}^{n} & C_{22}^{1}C_{22}^{n}
\end{array}\right)\left(\begin{array}{c}
A_{q}^{++}\\
A_{q}^{+-}\\
A_{q}^{-+}\\
A_{q}^{--}
\end{array}\right).\label{eq:amplitudes relations}
\end{equation}
In the above, $A_{q}^{++}$ is shorthand notation of $A_{q}(1_{Q}^{+},\cdots,n_{\bar{Q}}^{+})$
and all the amplitudes have the same external gluons. As in equation
(\ref{eq:spinor c xishu}), the elements of the matrix are 
\begin{eqnarray*}
C_{11}^{1} & = & \frac{\langle\tilde{q}|\slashed{p}_{1}|q]}{\langle\tilde{q}\tilde{p_{1}}^{\flat}\rangle[p_{1}^{\flat}q]},C_{12}^{1}=\frac{m\langle\tilde{q}q\rangle}{\langle\tilde{q}\tilde{p_{1}}^{\flat}\rangle\langle p_{1}^{\flat}q\rangle},C_{21}^{1}=\frac{m[\tilde{q}q]}{[\tilde{q}\tilde{p_{1}}^{\flat}][p_{1}^{\flat}q]},C_{22}^{1}=\frac{[\tilde{q}|\slashed{p}_{1}|q\rangle}{[\tilde{q}\tilde{p_{1}}^{\flat}]\langle p_{1}^{\flat}q\rangle},
\end{eqnarray*}
\begin{equation}
C_{11}^{n}=\frac{\langle\tilde{q}|\slashed{p}_{n}|q]}{\langle\tilde{q}\tilde{p_{n}}^{\flat}\rangle[p_{n}^{\flat}q]},C_{12}^{n}=\frac{m\langle\tilde{q}q\rangle}{\langle\tilde{q}\tilde{p_{n}}^{\flat}\rangle\langle p_{n}^{\flat}q\rangle},C_{21}^{n}=\frac{m[\tilde{q}q]}{[\tilde{q}\tilde{p_{n}}^{\flat}][p_{n}^{\flat}q]\,},C_{22}^{n}=\frac{[\tilde{q}|\slashed{p}_{n}|q\rangle}{[\tilde{q}\tilde{p_{n}}^{\flat}]\langle p_{n}^{\flat}q\rangle}.\label{eq:ampCxishu}
\end{equation}
These relations between massive amplitudes are important for our calculations
in the next section and the correctness of it will also be checked
there.

\section{calculation of fermion-pair amplitudes}

In this section, we use the onshell BCFW recursion relations to calculate
several $n$-point amplitudes with a pair of massive fermions and
$n-2$ gluons. First we list several excellent and useful results
about amplitudes with a pair of massive particles, which are the building
blocks of our calculation.

An excellent and compact expression for the amplitude of a massive
complex scalar-antiscalar pair and any number of positive helicity
gluons is obtained in \cite{Ferrario:2006np}: 
\begin{equation}
A(1_{\phi},2^{+},\cdots,(n-1)^{+},n_{\bar{\phi}})=2^{n/2-1}im^{2}\frac{[2|\prod_{k=3}^{n-2}(y_{1,k}-\slashed{p}_{k}\slashed{p}_{1,k-1})|n-1]}{y_{1,2}y_{1,3}\cdots y_{1,n-2}\langle23\rangle\langle34\rangle\cdots\langle n-2,n-1\rangle},\label{eq:scalarallplus}
\end{equation}
where 
\begin{eqnarray}
p_{1,k} & = & p_{1}+p_{2}+\cdots+p_{k},\\
y_{1,k} & = & (p_{1}+p_{2}+\cdots+p_{k})^{2}-m^{2}.
\end{eqnarray}

In supersymmetric theories, SUSY transformation can relate a bosonic
particle state to a fermionic particle state. And there are also relations
between amplitudes related by SUSY transformations. The SUSY transformations
of helicity states have been discussed in \cite{Grisaru:1976vm,Grisaru:1977px}.
In massless SUSY QCD theory, these SUSY transformations have been
applied to the helicity amplitudes \cite{Parke:1985pn,Kunszt:1985mg,Reuter:2002gn,Bidder:2005in}.
In massive SUSY QCD, similar transformations of helicity states have
been derived in \cite{schwinn0602}. Using SUSY transformations, some
relations between amplitudes with different external helicity particles
can also be obtained. These relations are the so-called SUSY Ward
identities\cite{schwinn0602}. As already mentioned, in this paper,
we choose the same reference momentum for the external massive fermions
of an amplitude. In this case, amplitudes of a massive fermion pair
and all plus helicity gluons have the following relations with amplitudes
where the fermion pair is replaced by massive scalar pair,
\begin{eqnarray}
A_{q}(1_{Q}^{+},2^{+},\cdots,(n-1)^{+},n_{\bar{Q}}^{-}) & = & \frac{\langle p_{n}^{\flat}q\rangle}{\langle p_{1}^{\flat}q\rangle}A(1_{\phi}^{+},2^{+},\cdots,(n-1)^{+},n_{\bar{\phi}}^{-}),\nonumber \\
A_{q}(1_{Q}^{-},2^{+},\cdots,(n-1)^{+},n_{\bar{Q}}^{+}) & = & -\frac{\langle p_{1}^{\flat}q\rangle}{\langle p_{n}^{\flat}q\rangle}A(1_{\phi}^{-},2^{+},\cdots,(n-1)^{+},n_{\bar{\phi}}^{+}),\nonumber \\
A_{q}(1_{Q}^{-},2^{+},\cdots,(n-1)^{+},n_{\bar{Q}}^{-}) & = & \frac{\langle p_{1}^{\flat}p_{n}^{\flat}\rangle}{m}A(1_{\phi}^{-},2^{+},\cdots,(n-1)^{+},n_{\bar{\phi}}^{-}).\label{eq:allplusrelation}
\end{eqnarray}
It is noted that all the SUSY Ward identities as the above are derived
by using supersymmetry and these relations should hold to any perturbative
orders. At the tree level some relations can be applied to non-supersymmetric
theory. This results from the fact that at tree level the quark-gluon
amplitudes are the same in both SUSY and non-SUSY theories and at
loop level there are contributions from SUSY particles to amplitudes.
The helicity flipping amplitude with all plus external particles vanish,
$A_{q}(1_{Q}^{+},2^{+},\cdots,(n-1)^{+},n_{\bar{Q}}^{+})=0$. Plugging
eq.(\ref{eq:scalarallplus}) into eq.(\ref{eq:allplusrelation}),
we can get the basic building blocks for the calculation of massive
fermion-pair amplitudes. 

There are also similar relations between fermion-pair amplitudes and
scalar-pair amplitudes with one minus helicity gluon
\begin{eqnarray}
A_{j}(1_{Q}^{+},2^{+},\cdots,j^{-},\cdots,(n-1)^{+},n_{\bar{Q}}^{-}) & = & \frac{\langle n_{j}j\rangle}{\langle1_{j}j\rangle}A(1_{\phi}^{+},2^{+},\cdots,j^{-},\cdots,(n-1)^{+},n_{\bar{\phi}}^{-}),\nonumber \\
A_{j}(1_{Q}^{-},2^{+},\cdots,j^{-},\cdots,(n-1)^{+},n_{\bar{Q}}^{+}) & = & -\frac{\langle1_{j}j\rangle}{\langle n_{j}j\rangle}A(1_{\phi}^{-},2^{+},\cdots,j^{-},\cdots,(n-1)^{+},n_{\bar{\phi}}^{+}),\label{eq:oneminus}
\end{eqnarray}
where $1_{j}=p_{1}^{\flat}|_{r=p_{j}}$ is the projection of $p_{1}$
when we choose $p_{j}$ as reference momentum. One can note that there
is not similar relations for helicity flipping amplitudes when there
are one minus helicity gluon.

Before we precede to the calculation, here we use the SUSY Ward identities
(\ref{eq:allplusrelation}) to check the correctness of the transformation
rules (\ref{eq:amplitudes relations}). We know that $A_{q}(1_{Q}^{+},2^{+},\cdots,(n-1)^{+},n_{\bar{Q}}^{+})=0$
is correct for any reference momentum $q$. Then using the transformation
rules (\ref{eq:amplitudes relations}), we can obtain
\begin{eqnarray}
A_{\tilde{q}}(1_{Q}^{+},2^{+},\cdots,(n-1)^{+},n_{\bar{Q}}^{+}) & = & C_{11}^{1}C_{11}^{n}A_{q}^{++}+C_{11}^{1}C_{12}^{n}A_{q}^{+-}+C_{12}^{1}C_{11}^{n}A_{q}^{-+}+C_{12}^{1}C_{12}^{n}A_{q}^{--}\nonumber \\
 & = & C_{11}^{1}C_{12}^{n}A_{q}^{+-}+C_{12}^{1}C_{11}^{n}A_{q}^{-+}+C_{12}^{1}C_{12}^{n}A_{q}^{--}\nonumber \\
 & = & (C_{11}^{1}C_{12}^{n}\frac{\langle nq\rangle}{\left\langle 1q\right\rangle }-C_{12}^{1}C_{11}^{n}\frac{\left\langle 1q\right\rangle }{\left\langle nq\right\rangle }+C_{12}^{1}C_{12}^{n}\frac{\left\langle 1n\right\rangle }{m})A(\phi),
\end{eqnarray}
where $A(\phi)=A(1_{\phi}^{+},2^{+},\cdots,(n-1)^{+},n_{\bar{\phi}}^{-})$
. The coefficient of $A(\phi)$ can be calculated as
\begin{eqnarray}
 &  & \frac{\langle\tilde{q}|\slashed{p}_{1}|q]}{\left\langle \tilde{q}\tilde{1}\right\rangle \left[1q\right]}\frac{m\left\langle \tilde{q}q\right\rangle }{\left\langle \tilde{q}\tilde{n}\right\rangle \left\langle nq\right\rangle }\frac{\left\langle nq\right\rangle }{\left\langle 1q\right\rangle }-\frac{\langle\tilde{q}|\slashed{p}_{n}|q]}{\left\langle \tilde{q}\tilde{n}\right\rangle \left[nq\right]}\frac{m\left\langle \tilde{q}q\right\rangle }{\left\langle \tilde{q}\tilde{1}\right\rangle \left\langle 1q\right\rangle }\frac{\left\langle 1q\right\rangle }{\left\langle nq\right\rangle }+\frac{m\left\langle \tilde{q}q\right\rangle }{\left\langle \tilde{q}\tilde{1}\right\rangle \left\langle 1q\right\rangle }\frac{m\left\langle \tilde{q}q\right\rangle }{\left\langle \tilde{q}\tilde{n}\right\rangle \left\langle nq\right\rangle }\frac{\left\langle 1n\right\rangle }{m}\nonumber \\
 & = & \frac{\langle\tilde{q}1\rangle}{\left\langle \tilde{q}\tilde{1}\right\rangle }\frac{m\left\langle \tilde{q}q\right\rangle }{\left\langle \tilde{q}\tilde{n}\right\rangle \left\langle nq\right\rangle }\frac{\left\langle nq\right\rangle }{\left\langle 1q\right\rangle }-\frac{\langle\tilde{q}n\rangle}{\left\langle \tilde{q}\tilde{n}\right\rangle }\frac{m\left\langle \tilde{q}q\right\rangle }{\left\langle \tilde{q}\tilde{1}\right\rangle \left\langle 1q\right\rangle }\frac{\left\langle 1q\right\rangle }{\left\langle nq\right\rangle }+\frac{\left\langle 1n\right\rangle \left\langle \tilde{q}q\right\rangle }{\left\langle \tilde{q}\tilde{1}\right\rangle \left\langle 1q\right\rangle }\frac{m\left\langle \tilde{q}q\right\rangle }{\left\langle \tilde{q}\tilde{n}\right\rangle \left\langle nq\right\rangle }\nonumber \\
 & = & \frac{m\left\langle \tilde{q}q\right\rangle }{\left\langle \tilde{q}\tilde{1}\right\rangle \left\langle \tilde{q}\tilde{n}\right\rangle \left\langle nq\right\rangle \left\langle 1q\right\rangle }(\langle\tilde{q}1\rangle\left\langle nq\right\rangle -\langle\tilde{q}n\rangle\left\langle 1q\right\rangle +\left\langle 1n\right\rangle \left\langle \tilde{q}q\right\rangle )\nonumber \\
 & = & 0.
\end{eqnarray}
So from the transformation rules, we can obtain the correct results
$A_{\tilde{q}}(\phi)=0$. In a similar way, one can check the correctness
of the transformation rules (\ref{eq:amplitudes relations}) from
other SUSY Ward identities.

From the SUSY Ward identities, we have obtained the fermion-pair amplitudes
with all plus helicity gluons. Then we can use onshell BCFW recursion
relation to get fermion-pair amplitudes with other gluon helicity
configurations. In \cite{Schwinn:2007ee}, it has been proved that
by choosing a proper momenta shift, we can always use onshell BCFW
recursion relation to calculate fermion-pair amplitudes. In the following,
we will calculate several concrete multigluon amplitudes.

The amplitude $A(1_{Q}^{+},2^{+},\cdots,(n-1)^{-},n_{\bar{Q}}^{-})$
can be calculated by shifting momenta $p_{n-1},p_{n}$. In spinor
formalism, it is 
\begin{eqnarray}
|n\rangle & \rightarrow & |n\rangle+z|n-1\rangle,\nonumber \\
|n-1] & \rightarrow & |n-1]-z|n],
\end{eqnarray}
where 
\begin{equation}
|n\pm\rangle=|p_{n}^{\flat}\pm\rangle,
\end{equation}
and
\begin{equation}
p_{n}=p_{n}^{\flat}+\frac{m^{2}}{2p_{n-1}\cdot p_{n}}p_{n-1}.
\end{equation}
Then using BCFW recursion relation, we can get a compact result as
follows, 
\begin{eqnarray}
 &  & A_{n-1}(1_{Q}^{+},2^{+},\cdots,(n-1)^{-},n_{\bar{Q}}^{-})\nonumber \\
 & = & \sum_{k=2}^{n-2}A_{n-1}(1_{Q}^{+},\cdots,(k-1)^{+},\hat{P}_{k,n-1}^{+},\hat{n}_{\bar{Q}}^{-})\frac{i}{p_{k,n-1}^{2}}A(k^{+},\cdots,\widehat{n-1}^{-},-\hat{P}_{k,n-1}^{-})\nonumber \\
 & = & i2^{n/2-1}\frac{\left\langle n,n-1\right\rangle }{\left\langle 1,n-1\right\rangle \left\langle 23\right\rangle \cdots\left\langle n-2,n-1\right\rangle }\sum_{k=2}^{n-2}\frac{\langle n-1|\slashed{p}_{k,n-1}\slashed{p}_{n}|n-1\rangle^{2}}{p_{k,n-1}^{2}\langle k|\slashed{p}_{k,n-1}\slashed{p}_{n}|n-1\rangle}\times\nonumber \\
 &  & \left(\delta_{k,2}+\delta_{k\neq2}\frac{m^{2}\left\langle k-1,k\right\rangle [2|\prod_{j=3}^{k-1}(y_{1,j}-\slashed{p}_{j}\slashed{p}_{1,j-1})\slashed{p}_{k,n-1}|n-1\rangle}{y_{1,2}\cdots y_{1,k-1}\langle k-1|\slashed{p}_{k,n-1}\slashed{p}_{n}|n-1\rangle}\right),\label{eq:n-1minus}
\end{eqnarray}
where $\delta_{k\neq2}=1-\delta_{k,2}$, and when $k=3$, $\prod_{j=3}^{k-1}(\cdots)=1$.
From the SUSY Ward identities, if we miss the factor $\frac{\left\langle n,n-1\right\rangle }{\left\langle 1,n-1\right\rangle }$,
we get the corresponding multigluon amplitude with massive scalar.
It has a more compact form than the one obtained in \cite{Forde:2005ue}
because we use more compact amplitudes eq.(\ref{eq:scalarallplus})
as building blocks. But we can check some lower point amplitudes with
others. The four point scalar amplitude is
\begin{equation}
A(1_{\phi}^{+},2^{+},3^{-},4_{\phi}^{-})=i2\frac{\langle3|\slashed{p}_{2,3}\slashed{p}_{4}|3\rangle^{2}}{\left\langle 23\right\rangle p_{2,3}^{2}\langle2|\slashed{p}_{2,3}\slashed{p}_{4}|3\rangle}=i2\frac{\langle3|\slashed{p}_{4}|2]}{p_{2,3}^{2}y_{1,2}}.
\end{equation}

The five point amplitude is 
\begin{eqnarray}
A(1_{\phi}^{+},2^{+},3^{+},4^{-},5_{\phi}^{-}) & = & i2^{3/2}\frac{\langle4|\slashed{p}_{2,4}\slashed{p}_{5}|4\rangle^{2}}{\left\langle 23\right\rangle \left\langle 34\right\rangle p_{2,4}^{2}\langle2|\slashed{p}_{2,4}\slashed{p}_{5}|4\rangle}\nonumber \\
 & + & i2^{3/2}\frac{m^{2}\langle4|\slashed{p}_{3,4}\slashed{p}_{5}|4\rangle^{2}}{\left\langle 34\right\rangle p_{3,4}^{2}\langle3|\slashed{p}_{3,4}\slashed{p}_{5}|4\rangle}\frac{[2|\slashed{p}_{3,4}|4\rangle}{y_{1,2}\langle2|\slashed{p}_{3,4}\slashed{p}_{5}|4\rangle}\nonumber \\
 & = & i2^{3/2}\frac{\langle4|\slashed{p}_{1}\slashed{p}_{2,4}|4\rangle^{2}}{\left\langle 23\right\rangle \left\langle 34\right\rangle p_{2,4}^{2}\langle2|\slashed{p}_{1}\slashed{p}_{2,4}|4\rangle}+i2^{3/2}\frac{m^{2}[3|\slashed{p}_{5}|4\rangle^{2}[23]}{[34]y_{1,2}y_{1,3}\langle4|\slashed{p}_{5}\slashed{p}_{3,4}|2\rangle}
\end{eqnarray}
These results are the same as the ones obtained from other ways \cite{Badger2005,Forde:2005ue}
up to overall conventional coefficients.

Then we calculate another four multigluon amplitudes with a massive
fermion-pair. For amplitude
\[
A(1_{Q}^{+},2^{+},3^{-},\cdots,(n-1)^{+},n_{\bar{Q}}^{+}),
\]
 we shift the momenta $p_{1}$ and $p_{3}$,
\begin{eqnarray}
|1\rangle & \rightarrow & |1\rangle+z|3\rangle,\nonumber \\
|3] & \rightarrow & |3]-z|1].\label{eq:A+--shift}
\end{eqnarray}
The amplitude $A(1_{Q}^{+},2^{+},3^{-},\cdots,(n-1)^{+},n_{\bar{Q}}^{+})$
can be decomposed as

\begin{eqnarray}
A_{3}(1_{Q}^{+},2^{+},3^{-},\cdots,(n-1)^{+},n_{\bar{Q}}^{+}) & = & \sum_{k=4}^{n-1}A_{3}(\hat{1}_{Q}^{+},2^{+},\hat{P}_{3,k}^{+},\cdots,n_{\bar{Q}}^{+})\frac{i}{p_{3,k}^{2}}A(-\hat{P}_{3,k}^{-},\hat{3}^{-},\cdots,k^{+})\nonumber \\
 & + & \sum_{l=3}^{n-1}A_{3}(\hat{1}_{Q}^{+},2^{+},\hat{P}_{2,l}^{+},\cdots,n_{\bar{Q}}^{+})\frac{i}{p_{2,l}^{2}}A(-\hat{P}_{2,l}^{-},2^{+},\hat{3}^{-},\cdots,l^{+})\nonumber \\
 & + & A_{3}(\hat{1}_{Q}^{+},2^{+},-\hat{P}_{12}^{-})\frac{1}{p_{12}^{2}}A_{3}(\hat{P}_{12}^{+},\hat{3}^{-},\cdots,n_{\bar{Q}}^{+}).\label{eq:A+-+}
\end{eqnarray}
It is easy to see the first two terms in the above equation are both
zero because there are fermion-pair amplitudes with all plus helicity.
Let us see the third term $A_{3}(\hat{1}_{Q}^{+},2^{+},-\hat{P}_{12}^{-})\frac{1}{p_{12}^{2}}A_{3}(\hat{P}_{12}^{+},\hat{3}^{-},\cdots,n_{\bar{Q}}^{+})$.
We already know 
\begin{equation}
A_{\hat{3}}(\hat{P}_{12}^{+},\hat{3}^{-},\cdots,n_{\bar{Q}}^{+})=0.
\end{equation}
Using the transformation of amplitude with respect to different momenta,
it is easy to show 
\begin{equation}
A_{3}(\hat{P}_{12}^{+},\hat{3}^{-},\cdots,n_{\bar{Q}}^{+})=0.
\end{equation}
Then 
\begin{equation}
A_{3}(1_{Q}^{+},2^{+},3^{-},\cdots,(n-1)^{+},n_{\bar{Q}}^{+})=0.
\end{equation}
Using the same recursive calculation and induction, we can prove

\begin{equation}
A_{j}(1_{Q}^{+},2^{+},\cdots,j^{-},\cdots,(n-1)^{+},n_{\bar{Q}}^{+})=0.
\end{equation}
This is consistent with the result from SUSY Ward identities\cite{schwinn0602},
which are obtained only by the supersymmetry of massive SUSY QCD.

Then we use the same shifting of momenta as eq.(\ref{eq:A+--shift})
and calculate the amplitude $A_{3}(1_{Q}^{+},2^{+},3^{-},\cdots,(n-1)^{+},n_{\bar{Q}}^{-})$.
We obtain 

\begin{eqnarray}\label{A+-} &&A_{3}(1_{Q}^{+},2^{+},3^{-},\cdots,(n-1)^{+},n_{\bar{Q}}^{-})
\\\nonumber
&=&i2^{n/2-1}\frac{\left\langle n3\right\rangle}{\left\langle 13\right\rangle }\frac{m^{2}}
{\left\langle 34\right\rangle \cdots\left\langle n-2,n-1\right\rangle }\\\nonumber
&\times&\sum_{k=4}^{n-1}  
\frac{ \langle3|\slashed{p}_{1}\slashed{p}_{3,k}|3\rangle^{3}}{p_{3,k}^{2}\langle3|\slashed{p}_{1}\slashed{p}_{3,k}|2\rangle\langle3|\slashed{p}_{1}\slashed{p}_{3,k}|k\rangle
(\langle 3|\slashed{p}_{1}\slashed{p}_{3,k}|3\rangle y_{1,2}
+\langle 3|\slashed{p}_{1}\slashed{p}_{1,2}|3 \rangle p_{3,k}^2) }\\\nonumber 
&\times&\{\frac{\delta_{k\neq n-1}\left\langle k,k+1\right\rangle
\langle3|\slashed{p}_{1}\slashed{p}_{3,k}|3\rangle} 
{y_{1,k}\cdots y_{1,n-2}
\langle3|\slashed{p}_{1}\slashed{p}_{3,k}|k+1\rangle
}
([2|(y_{1,k}+\slashed{p}_{3,k}\slashed{p}_{k+1,n})\prod_{j=k+1}^{n-2}(y_{1,j}-\slashed{p}_{j}\slashed{p}_{1,j-1})|n-1]\\\nonumber
&+&\frac{\langle 3|\slashed{p}_1|2]} 
{\langle3|\slashed{p}_{1}\slashed{p}_{3,k}|3\rangle}
p_{3,k}^2\langle 3|\slashed{p}_{k+1,n}
\prod_{j=k+1}^{n-2}(y_{1,j}-\slashed{p}_{j}\slashed{p}_{1,j-1})|n-1])+
\delta_{k,n-1}\langle 3|\slashed{p}_{3,n-1}|2]  \}\\\nonumber
&+&i2^{n/2-1}\frac{\left\langle n3\right\rangle}{\left\langle 13\right\rangle }\frac{1}{
\langle  23\rangle\cdots\langle  n-2,n-1\rangle } 
\sum_{l=3}^{n-1}
\frac{\langle 3|\slashed{p}_{1}\slashed{p}_{2,l}|3\rangle^3}
{p_{2,l}^2
\langle 3|\slashed{p}_{1}\slashed{p}_{2,l}|2\rangle
\langle 3|\slashed{p}_{1}\slashed{p}_{2,l}|l\rangle}\\\nonumber
&\times&
\{\delta_{l,n-1}+\delta_{l\neq n-1}m^{2}\frac{\langle l, l+1 \rangle
\langle 3|\slashed{p}_{2,l}\prod_{j=l+1}^{n-2}(y_{1,j}-\slashed{p}_{j}\slashed{p}_{1,j-1})|n-1]}{ y_{1,l}\cdots y_{1,n-2}
\langle 3|\slashed{p}_{1}\slashed{p}_{2,l}|l+1\rangle}
\}\\\nonumber
&+&i2^{n/2-1}\frac{\left\langle n3\right \rangle}{\left\langle 13\right\rangle }
\frac{\langle 3|\slashed{p}_1|2]}{y_{1,2}\langle 23\rangle\cdots\langle 
n-2,n-1\rangle}\\\nonumber
&\times&\sum_{j=4}^{n-1}\frac{
\langle 3|\slashed{p}_{1}\slashed{p}_{2}|3\rangle
\langle 3|\slashed{p}_{1,2}\slashed{p}_{3,j}|3\rangle^2}
{(\langle 3|\slashed{p}_{1}\slashed{p}_{2}|3\rangle
p_{3,j}^2+
\langle 3|\slashed{p}_{1}\slashed{p}_{3,j}|3\rangle y_{1,2})
\langle 3|(y_{1,2}+\slashed{p}_{1,2}\slashed{p}_{3,j})|j\rangle}
\\\nonumber
&\times&\{\delta_{j,n-1}+\delta_{j\neq n-1}m^2\frac{\langle j,j+1\rangle
\langle 3 |\slashed{p}_{3,j}\prod_{k=j+1}^{n-2}(y_{1,k}-\slashed{p}_{k}\slashed{p}_{1,k-1})|n-1]}
{y_{1,j}\cdots y_{1,n-2}
\langle 3|(y_{1,2}+\slashed{p}_{1,2}\slashed{p}_{3,j})|j+1\rangle
}\}.
\end{eqnarray}

For multigluon amplitude $A_{3}(1_{Q}^{-},2^{+},3^{-},\cdots,(n-1)^{+},n_{\bar{Q}}^{+})$,
we can use the similar recursive method to calculate and we obtain

\begin{eqnarray}\label{A-+} &&A_{3}(1_{Q}^{-},2^{+},3^{-},\cdots,(n-1)^{+},n_{\bar{Q}}^{+})
\\\nonumber
&=&-i2^{n/2-1}\frac{\left\langle 13\right\rangle}{\left\langle n3\right\rangle }\frac{m^{2}}
{\left\langle 34\right\rangle \cdots\left\langle n-2,n-1\right\rangle }\\\nonumber
&\times&\sum_{k=4}^{n-1}  
\frac{ \langle3|\slashed{p}_{1}\slashed{p}_{3,k}|3\rangle^{3}}{p_{3,k}^{2}\langle3|\slashed{p}_{1}\slashed{p}_{3,k}|2\rangle\langle3|\slashed{p}_{1}\slashed{p}_{3,k}|k\rangle
(\langle 3|\slashed{p}_{1}\slashed{p}_{3,k}|3\rangle y_{1,2}
+\langle 3|\slashed{p}_{1}\slashed{p}_{1,2}|3 \rangle p_{3,k}^2) }\\\nonumber 
&\times&\{\frac{\delta_{k\neq n-1}\left\langle k,k+1\right\rangle
\langle3|\slashed{p}_{1}\slashed{p}_{3,k}|3\rangle} 
{y_{1,k}\cdots y_{1,n-2}
\langle3|\slashed{p}_{1}\slashed{p}_{3,k}|k+1\rangle
}
([2|(y_{1,k}+\slashed{p}_{3,k}\slashed{p}_{k+1,n})\prod_{j=k+1}^{n-2}(y_{1,j}-\slashed{p}_{j}\slashed{p}_{1,j-1})|n-1]\\\nonumber
&+&\frac{\langle 3|\slashed{p}_1|2]} 
{\langle3|\slashed{p}_{1}\slashed{p}_{3,k}|3\rangle}
p_{3,k}^2\langle 3|\slashed{p}_{k+1,n}
\prod_{j=k+1}^{n-2}(y_{1,j}-\slashed{p}_{j}\slashed{p}_{1,j-1})|n-1])+
\delta_{k,n-1}\langle 3|\slashed{p}_{3,n-1}|2]  \}\\\nonumber
&-&i2^{n/2-1}\frac{\left\langle 13\right\rangle}{\left\langle n3\right\rangle }\frac{1}{
\langle  23\rangle\cdots\langle  n-2,n-1\rangle } 
\sum_{l=3}^{n-1}
\frac{\langle 3|\slashed{p}_{1}\slashed{p}_{2,l}|3\rangle^3}
{p_{2,l}^2
\langle 3|\slashed{p}_{1}\slashed{p}_{2,l}|2\rangle
\langle 3|\slashed{p}_{1}\slashed{p}_{2,l}|l\rangle}\\\nonumber
&\times&
\{\delta_{l,n-1}+\delta_{l\neq n-1}m^{2}\frac{\langle l, l+1 \rangle
\langle 3|\slashed{p}_{2,l}\prod_{j=l+1}^{n-2}(y_{1,j}-\slashed{p}_{j}\slashed{p}_{1,j-1})|n-1]}{ y_{1,l}\cdots y_{1,n-2}
\langle 3|\slashed{p}_{1}\slashed{p}_{2,l}|l+1\rangle}
\}\\\nonumber
&-&i2^{n/2-1}\frac{\left\langle 13\right \rangle}{\left\langle n3\right\rangle }
\frac{\langle 3|\slashed{p}_1|2]}{y_{1,2}\langle 23\rangle\cdots\langle 
n-2,n-1\rangle}\\\nonumber
&\times&\sum_{j=4}^{n-1}\frac{
\langle 3|\slashed{p}_{1}\slashed{p}_{2}|3\rangle
\langle 3|\slashed{p}_{1,2}\slashed{p}_{3,j}|3\rangle^2}
{(\langle 3|\slashed{p}_{1}\slashed{p}_{2}|3\rangle
p_{3,j}^2+
\langle 3|\slashed{p}_{1}\slashed{p}_{3,j}|3\rangle y_{1,2})
\langle 3|(y_{1,2}+\slashed{p}_{1,2}\slashed{p}_{3,j})|j\rangle}
\\\nonumber
&\times&\{\delta_{j,n-1}+\delta_{j\neq n-1}m^2\frac{\langle j,j+1\rangle
\langle 3 |\slashed{p}_{3,j}\prod_{k=j+1}^{n-2}(y_{1,k}-\slashed{p}_{k}\slashed{p}_{1,k-1})|n-1]}
{y_{1,j}\cdots y_{1,n-2}
\langle 3|(y_{1,2}+\slashed{p}_{1,2}\slashed{p}_{3,j})|j+1\rangle
}\}.
\end{eqnarray}

Compairing this result with eq.(\ref{A+-}), we can see that they
are different from each other just by a constant coefficient. And
this is consistent with the SUSY Ward identities in eq.(\ref{eq:oneminus}).
Just as eq.(\ref{eq:n-1minus}), eq.(\ref{A+-}) and eq.(\ref{A-+})
have more compact forms than the ones obtained from other ways. 

Then let us check the massless limit of eq.(\ref{A+-}). Taking $m=0$,
eq.(\ref{A+-}) becomes\begin{eqnarray}\label{A+-massless} &&A_{3}(1_{Q}^{+},2^{+},3^{-},\cdots,(n-1)^{+},n_{\bar{Q}}^{-})
\\\nonumber
&=&i2^{n/2-1}\frac{\left\langle n3\right\rangle}{\left\langle 13\right\rangle }\frac{1}{
\langle  23\rangle\cdots\langle  n-2,n-1\rangle } 
\frac{\langle 3|\slashed{p}_{n-1}\slashed{p}_{2,n-1}|3\rangle^3}
{p_{2,n-1}^2
\langle 3|\slashed{p}_{n-1}\slashed{p}_{2,n-1}|2\rangle
\langle 3|\slashed{p}_{n-1}\slashed{p}_{2,n-1}|n-1\rangle}\\\nonumber
&+&i2^{n/2-1}\frac{\left\langle n3\right \rangle}{\left\langle 13\right\rangle }
\frac{\langle 3|\slashed{p}_1|2]}{y_{1,2}\langle 23\rangle\cdots\langle 
n-2,n-1\rangle}\\\nonumber
&\times&\frac{
\langle 3|\slashed{p}_{1}\slashed{p}_{2}|3\rangle
\langle 3|\slashed{p}_{1,2}\slashed{p}_{3,n-1}|3\rangle^2}
{(\langle 3|\slashed{p}_{1}\slashed{p}_{2}|3\rangle
p_{3,n-1}^2+
\langle 3|\slashed{p}_{1}\slashed{p}_{3,n-1}|3\rangle y_{1,2})
\langle 3|(y_{1,2}+\slashed{p}_{1,2}\slashed{p}_{3,n-1})|n-1\rangle}\\\nonumber
&=&i2^{n/2-1}\frac{\langle n3\rangle^3\langle 31\rangle}
{\langle 12\rangle\cdots\langle n1\rangle}.
\end{eqnarray}This is exactly the MHV amplitude with a fermion-antifermion pair
in massless QCD theory. 

Finally, we calculate the amplitude $A_{3}(1_{Q}^{-},2^{+},3^{-},\cdots,(n-1)^{+},n_{\bar{Q}}^{-})$
. There is no SUSY Ward identity to relate this amplitude with the
corresponding scalar one. So this kind of amplitudes should be calculated
directly from lower point amplitudes with a pair of fermions using
onshell recursive method. The result is 

\begin{eqnarray}\label{A--}
&&A_{3}(1_{Q}^{-},2^{+},3^{-},\cdots,(n-1)^{+},n_{\bar{Q}}^{-})\\ \nonumber
&=&\frac{i 2^{n/2-1}m}
{\left\langle 34\right\rangle \cdots\left\langle n-2,n-1\right\rangle}
\\\nonumber
&\times&\sum_{k=4}^{n-1}  
(\langle 1n \rangle+\frac{\langle 3n \rangle}
{\langle 3|\slashed{p}_{3,k}|1]}p_{3,k}^{2})
\frac{ \langle3|\slashed{p}_{1}\slashed{p}_{3,k}|3\rangle^{3}}{p_{3,k}^{2}\langle3|\slashed{p}_{1}\slashed{p}_{3,k}|2\rangle\langle3|\slashed{p}_{1}\slashed{p}_{3,k}|k\rangle
(\langle 3|\slashed{p}_{1}\slashed{p}_{3,k}|3\rangle y_{1,2}
+\langle 3|\slashed{p}_{1}\slashed{p}_{1,2}|3 \rangle p_{3,k}^2) }\\\nonumber  
&\times&\{\frac{\delta_{k\neq n-1}\left\langle k,k+1\right\rangle
\langle3|\slashed{p}_{1}\slashed{p}_{3,k}|3\rangle} 
{y_{1,k}\cdots y_{1,n-2}
\langle3|\slashed{p}_{1}\slashed{p}_{3,k}|k+1\rangle
}
([2|(y_{1,k}+\slashed{p}_{3,k}\slashed{p}_{k+1,n})\prod_{j=k+1}^{n-2}(y_{1,j}-\slashed{p}_{j}\slashed{p}_{1,j-1})|n-1]\\\nonumber
&+&\frac{\langle 3|\slashed{p}_1|2]} 
{\langle3|\slashed{p}_{1}\slashed{p}_{3,k}|3\rangle}
p_{3,k}^2\langle 3|\slashed{p}_{k+1,n}
\prod_{j=k+1}^{n-2}(y_{1,j}-\slashed{p}_{j}\slashed{p}_{1,j-1})|n-1])+
\delta_{k,n-1}\langle 3|\slashed{p}_{3,n-1}|2]  \}\\\nonumber
&+&\frac{i2^{n/2-1}}
{m\langle  23\rangle\cdots\langle  n-2,n-1\rangle } 
\sum_{l=3}^{n-1}
(\langle 1n \rangle+\frac{\langle 3n \rangle}
{\langle 3|\slashed{p}_{2,l}|1]}p_{2,l}^{2})
\frac{\langle 3|\slashed{p}_{1}\slashed{p}_{2,l}|3\rangle^3}
{p_{2,l}^2
\langle 3|\slashed{p}_{1}\slashed{p}_{2,l}|2\rangle
\langle 3|\slashed{p}_{1}\slashed{p}_{2,l}|l\rangle}\\\nonumber
&\times&
\{\delta_{l,n-1}+\delta_{l\neq n-1}m^{2}\frac{\langle l, l+1 \rangle
\langle 3|\slashed{p}_{2,l}\prod_{j=l+1}^{n-2}(y_{1,j}-\slashed{p}_{j}\slashed{p}_{1,j-1})|n-1]}{ y_{1,l}\cdots y_{1,n-2}
\langle 3|\slashed{p}_{1}\slashed{p}_{2,l}|l+1\rangle}
\}\\\nonumber
&+&\frac{i2^{n/2-1}}
{\langle 34\rangle\cdots\langle n-2,n-1\rangle}
\sum_{j=4}^{n-1}\frac{
\langle 3|\slashed{p}_{1}\slashed{p}_{2}|3\rangle
\langle 3|\slashed{p}_{1,2}\slashed{p}_{3,j}|3\rangle^2 C(j)}
{(\langle 3|\slashed{p}_{1}\slashed{p}_{2}|3\rangle
p_{3,j}^2+
\langle 3|\slashed{p}_{1}\slashed{p}_{3,j}|3\rangle y_{1,2})
\langle 3|(y_{1,2}+\slashed{p}_{1,2}\slashed{p}_{3,j})|j\rangle}
\\\nonumber
&\times&\{\delta_{j,n-1}+\delta_{j\neq n-1}m^2\frac{\langle j,j+1\rangle
\langle 3 |\slashed{p}_{3,j}\prod_{k=j+1}^{n-2}(y_{1,k}-\slashed{p}_{k}\slashed{p}_{1,k-1})|n-1]}
{y_{1,j}\cdots y_{1,n-2}
\langle 3|(y_{1,2}+\slashed{p}_{1,2}\slashed{p}_{3,j})|j+1\rangle
}\},
\end{eqnarray}where $C(j)$ is 

\begin{eqnarray}\label{xishu}
C(j)&=&\frac{m[23]\langle n|\slashed{p}_{3}|2]}
{y_{1,2}[31]\langle 3|\slashed{p}_{1,2}|3]}
-\frac{m \langle 13\rangle^2\langle n|\slashed{p}_{3}|1]}
{y_{1,3}\langle 23\rangle^2 \langle 3|\slashed{p}_{1,2}|3]}
-\frac{m\langle 13\rangle\langle 3|\slashed{p}_{1}|3]}
{\langle 23\rangle\langle 3|\slashed{p}_{n}|3]
\langle 2|(\slashed{p}_{1}^\flat+\slashed{p}_{3})|n]}\\\nonumber
&+&\frac{\langle 13\rangle\langle 3|\slashed{p}_{1}|2]}
{m\langle 23\rangle y_{1,2}}
(\frac{y_{1,3}\langle n|\slashed{p}_{1,2}|1]-
m^2\langle n|\slashed{p}_{3}|1]}
{y_{1,3}\langle 3|\slashed{p}_{1,2}|1]}-
\frac{y_{1,2}\langle n|\slashed{p}_{n}|1]}
{y_{1,2}\langle 3|\slashed{p}_{n}|1]
+2 p_3\cdot p_n\langle 3|\slashed{p}_{2}|1]})\\\nonumber
&+&\frac{\langle 13\rangle \langle n3\rangle 
\langle 3|\slashed{p}_{1}|2]}
{m\langle 23\rangle y_{1,2}
\langle 3|(y_{1,2}+\slashed{p}_{1,2}\slashed{p}_{3,j})|3\rangle}
(p_{3,j}^2+y_{1,2}
\frac{\langle 3|\slashed{p}_{3,j}|1]}{\langle 3|\slashed{p}_{2}|1]}).
\end{eqnarray}

In principle, for an $n$-point massive fermion-pair amplitude with
definite minus helicity gluon $j$ , we can use the above mentioned
shifting to get the analytic expression for it from amplitudes with
one minus helicity gluon nearer to the fermions. In fact, it is difficult
for doing it by hand except for some special gluon helicity configurations.
But because the shift of momenta, recursion decomposition and transformation
of amplitudes with respect to different reference momenta are all
systematic procedures, so it is suitable to develop a program to do
the work.

\section{summary }

In this paper, using onshell BCFW recursion relation and shifting
the momenta of a massive fermion and a gluon, we calculate several
tree level $n$-point amplitudes with a massive fermion-antifermion
pair and one minus helicity gluon. Amplitudes with massive fermions
depend on reference momentum for defining the helicity states of massive
external fermions, so it is more difficult for calculating them. We
give the explicit transformation rules for amplitudes with respect
to different reference momenta, which are important in the analytic
calculation of massive fermion-pair amplitudes. The correctness of
these rules are checked by SUSY Ward identities. We use the most compact
and excellent results for some special gluon helicity configurations,
such as all plus helicity, and pure gluon amplitudes as building blocks
to obtain four $n$-point massive amplitudes with more complicated
gluon helicity configurations. Generally, calculating the analytic
results of $n$-point massive amplitudes by hand is difficult. A program
need to be developed to calculate the amplitudes. But it is still
interesting and calculable to use the recursion method to explore
the analytic results of amplitudes with finite external particles
and more fermions, which are important in high energy physics experiments.
\begin{acknowledgments}
We would like to thank Professor Bo Feng for useful discussions and
reading the manuscript. This work is supported by Chinese NSF funding
under contract No. 11031005, No. 11125523. \end{acknowledgments}

\end{document}